\documentclass{moriond}

\usepackage{amsmath,amssymb,amsfonts}
\usepackage{bm}
\usepackage[T1]{fontenc}

\newcommand{\Photo}{}

\usepackage{mathtools}
\usepackage{bm}
\usepackage{feynmp}

\newcommand{\eqgraph}[3]{\vcenter{\hbox{\raisebox{#2}{#3}}}}

\begin{document}

\begin{fmffile}{diagrams}

\vspace*{4cm}

\title{Primordial black holes from inflation:
\\
on the decoupling between large and small scales}

\author{L. Iacconi}

\address{Astronomy Unit, Queen Mary University of London, London, E1 4NS, UK
\\
Institute of Cosmology and Gravitation, University of Portsmouth, Portsmouth, PO1 3FX, UK}

\maketitle\abstracts{
Primordial black holes (PBHs) can be produced from inflation if the primordial curvature power spectrum is strongly enhanced on scales much shorter than those probed by cosmic microwave background (CMB) experiments. 
In single-field models this typically requires a transient departure from slow-roll, attractor dynamics, for example realized through a brief ultra-slow-roll phase. 
In these scenarios, there is reasonable concern that large-scale modes, whose statistics is tightly constrained by CMB observations, might back-react to the amplified perturbations on much shorter scales. 
In a perturbative expansion for the long-mode power spectrum, this effect first appears at 1-loop. 
In these proceedings we summarize recent works on this issue, based on the application of the 
separate-universe framework and its general extension with multi-point propagators. 
We show that back-reaction at 1-loop is due to either (i) non-linear super-horizon evolution, or (ii) 1-loop-corrected initial conditions. 
By assuming separation of scales and adiabaticity of the long mode, we show that the 1-loop back-reaction is not observable and large scales decouple from enhanced short ones.
While we demonstrate that PBH production within single-field inflation does not disrupt large-scale predictions, we close by discussing scenarios to which our results do not apply.
}

\section{Introduction}

Primordial black holes (PBHs) offer a distinctive probe of the early Universe. 
In contrast to astrophysical BHs, they can in principle populate a continuous mass spectrum, and might be produced with sub-solar masses. 
In the context of cosmology, PBHs have been investigated, e.g., as dark-matter candidates, contributors to observed gravitational wave events and progenitors of super-massive black holes observed in the center of galaxies.
See Ref.~\cite{Carr:2026hot} for a recent summary of current observational constraints on PBHs and potential evidence for their existence. 

PBHs also offer a unique window on inflationary dynamics at scales much smaller than those probed by cosmic microwave background (CMB) experiments. 
Indeed, among other formation mechanisms, PBHs might be produced when rare, large-amplitude perturbations seeded during inflation re-enter the horizon during the standard hot Big Bang expansion and collapse gravitationally.
In this case, the abundance of formed PBHs is extremely sensitive to the statistics of the primordial curvature perturbation, $\zeta$, and observational limits on PBHs can be employed to constrain inflationary scenarios~\cite{Gow:2020bzo}.  
\vspace{0.3cm}

\noindent
\textbf{PBHs from large and rare primordial perturbations.}
A perturbation of characteristic comoving scale $k$ will collapse to form a PBH if, upon horizon re-entry ($k=aH$), its associated density contrast exceeds a threshold value, $\delta_c$.
By parametrising the perturbation with the value of the mass contained within the horizon at the time of collapse, $M_H$, the PBH abundance at formation can be written schematically as
\begin{equation}
\label{eq:beta}
    \beta(M_H) \approx \int_{\delta_c}^{\infty} \mathrm{d} \delta_{M_H} \, P(\delta_{M_H}) \;, 
\end{equation}
where $P(\delta_{M_H})$ denotes the probability distribution for the density contrast.
Eq.~\eqref{eq:beta} shows that PBHs are formed from large and rare perturbations, i.e. those that live in the tail of the distribution function.  

For the purpose of this qualitative discussion, let us assume that the probability distribution of the density contrast is Gaussian. 
Its variance, $\sigma(\delta_{M_H})$, is the typical size of the fluctuations, and it can be related to the primordial power spectrum as $\sigma^2(\delta_{M_H}) \sim \mathcal{P}_\zeta(k)$.
On CMB scales one has $\mathcal{P}_\zeta(k_{\mathrm{CMB}}) \sim 10^{-9}$, which is far too small to generate PBHs. 
By contrast, if $\mathcal{P}_\zeta(k_{\mathrm{PBH}}) \sim 10^{-3}$ on a much shorter scale, $k_{\mathrm{PBH}}$, then PBHs can be produced with appreciable abundance. 
\vspace{0.3cm}

\noindent
\textbf{PBHs from single-field inflation: the need for non-attractor dynamics.}
In order to produce PBHs from inflation it is necessary for the curvature perturbation to be strongly enhanced on scales smaller than those probed by the CMB. 
It is not possible to realize such a large amplification within standard single-field 
\footnote{While this can be achieved within multi-field dynamics, for the purpose of this discussion we focus on single-field models.}
slow-roll inflation, and a transient non-attractor phase embedded between two slow-roll eras is required. 
This can be understood by considering the super-horizon evolution of the curvature perturbation
\begin{equation}
\label{eq:superhorizon}
    \zeta_k(N)\big|_{k\ll aH} = c_1 + c_2 \int^N \mathrm{d} N' \, \exp\bigg[-\int^{N'} \mathrm{d} N''\, (3-\epsilon_1+\epsilon_2)\bigg] \;, 
\end{equation}
where time is measured in e-folds,  $\mathrm{d}N\equiv H \mathrm{d}t$, and the first two Hubble slow-roll parameters are defined as 
\begin{equation}
\label{eq:slowrollpars}
\epsilon_1 \equiv -\frac{\dot H}{H^2}\;,
\qquad
\epsilon_2 \equiv \frac{\dot \epsilon_1}{H \epsilon_1}\;. 
\end{equation}
During conventional slow-roll evolution $\epsilon_1,\epsilon_2 \ll 1$, and the second term in Eq.~\eqref{eq:superhorizon} decays exponentially, effectively freezing $\zeta_k$ on super-horizon. 
However, if there is no attractor in phase-space $(\epsilon_2<-3)$ the would-be decaying mode instead grows exponentially.
This super-horizon growth is the origin of the enhancement in the small-scale power spectrum.
A simple and widely studied example of non-attractor dynamics is ultra-slow-roll (USR)~\cite{Kinney:2005vj}, in which case $\epsilon_2\simeq -6$ transiently.
\begin{figure}
\begin{minipage}{0.4\linewidth}
\centerline{\includegraphics[width=\linewidth]{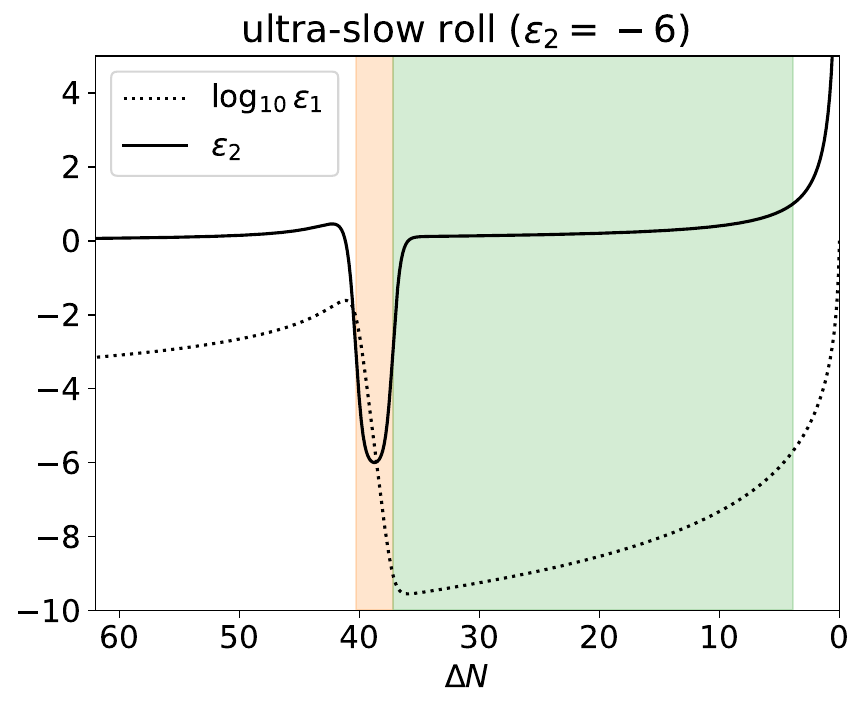}}
\end{minipage}
\hfill
\begin{minipage}{0.44\linewidth}
\centerline{\includegraphics[width=\linewidth]{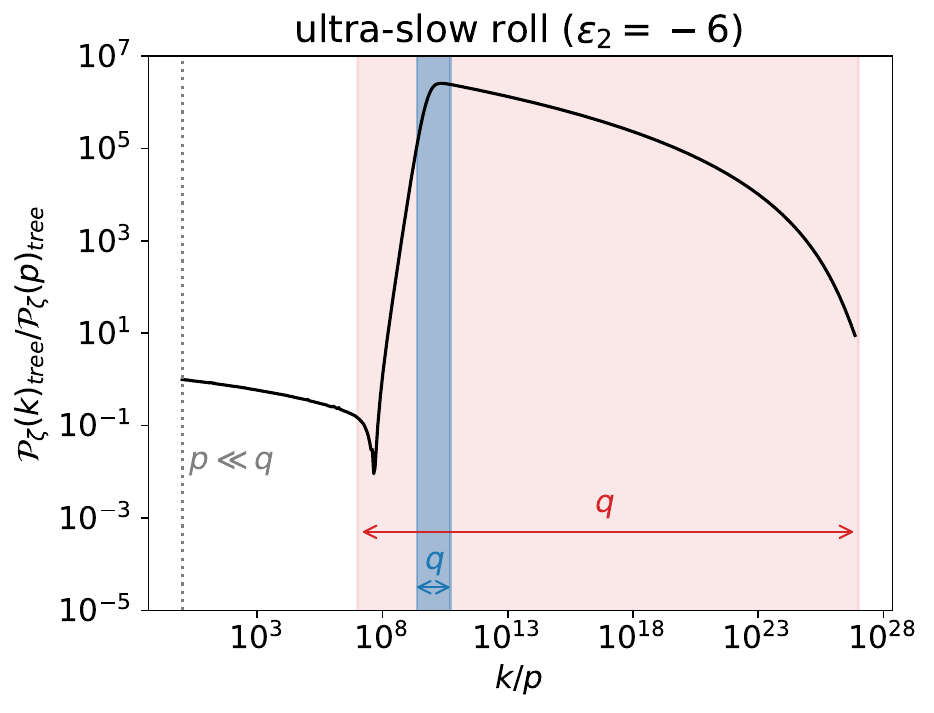}}
\end{minipage}
\caption[]{\textit{Left panel:} 
        Time-evolution of $\epsilon_1$ and $\epsilon_2$
        for a model featuring a transient USR phase. 
        The horizontal axis is
        $\Delta N \equiv N_\text{end}-N$,
        where $N_\text{end}$ labels the end of inflation. 
        We highlight times during the non-attractor phase
        $(\epsilon_2<-3)$ in orange,
        and the following attractor phase in green. 
        \textit{Right panel:} 
        Tree-level scalar power spectrum ${\mathcal{P}_\zeta(k)}_\text{tree}$,
        for the same USR model.
        The vertical dotted line identifies a long
        adiabatic mode, $p$. 
        Modes highlighted in blue cross the horizon during the non-attractor
        phase $(\epsilon_2<-3)$.
        We highlight the broad peak in red.}
\label{fig:USR}
\end{figure}
In Fig.~\ref{fig:USR} we represent the time-evolution of the first two slow-roll parameters and the resulting (tree-level) power spectrum for a USR model~\cite{Germani:2017bcs}. 
From now on we will label a large-scale, adiabatic mode compatible with CMB measurements as $p$, and modes belonging to the broadband of enhanced modes on short scales as $q$, with $p\ll q$.
\vspace{0.3cm}

\noindent
\textbf{Large-scale back-reaction.}
Tree-level slow-roll predictions for the large-scale statistics of primordial perturbations are in very good agreement with CMB observations~\cite{Planck:2018jri}.
Nevertheless, for any theory with interactions there will be loop contributions to the long-mode power spectrum. 
In models that can lead to PBH production, these inevitably include the effect of the amplified short-scale perturbations. 
There is therefore reasonable concern that significant back-reaction might occur, potentially disrupting the tree-level large-scale predictions. 
This was the motivation behind the analysis of Ref.~\cite{Kristiano:2022maq}, and many that followed. 
The authors of Ref.~\cite{Kristiano:2022maq} worked under the following set of simplifying assumptions that allowed them to carry out analytical computations: (i) the transitions between different phases of SR and USR evolution are modeled as instantaneous, (ii) in the loop integral short-scale modes are included up to the most enhanced one ($k_\text{peak}$), (iii) only 1-loop contributions generated by the insertion of two cubic interactions $H_\text{int}^{(3)}\propto \epsilon_2'$ (where a prime denotes a derivative with respect to time) were considered, (iv) UV modes that were still in the vacuum when inflation ended were not included. 
By performing a In--In computation, the large-scale back-reaction at 1-loop was found to be proportional to the peak amplitude, $\mathcal{P}_\zeta(p)_\text{1-loop}\propto \mathcal{P}_\zeta(k_\text{peak})_\text{tree}$, putting single-field PBH formation models in tension with perturbativity requirements~\cite{Kristiano:2022maq}. 
These results have been the subject of intense discussion. 
The purpose of these proceedings is to summarize the approach we proposed in Refs.~\cite{Iacconi:2023ggt,Iacconi:2026uzo}.
Thanks to the large separation of scales between the long and short modes, 
it is possible to compute the back-reaction at 1-loop within the separate universe framework~\cite{Wands:2000dp}, 
which provides a more transparent organization of the calculation than the In--In formalism.  
Moreover, we are able to go beyond some of the simplifying assumptions made in Ref.~\cite{Kristiano:2022maq}. 
In particular, we include 1-loop contributions from all cubic and quartic interactions, no assumption is made on the quality of the transition between the different dynamical phases, and we account for the effect of the \textit{whole} broadband of enhanced scales, including those falling away from the peak. 

\section{Back-reaction at 1-loop from separate universe}

The separate universe framework is a tool to describe the non-linear evolution of large-scale
cosmological perturbations. 
It is based on the observation that their evolution reduces to that of the background when gradients can be neglected.
At the end of inflation, a universe perturbed by a mode with comoving wavenumber $\mathbf{k}$ can therefore be described by sewing together uniform patches of size $L\ll k^{-1}$, each one initialised with slightly different initial conditions and evolved according to the background equations of motion. 
For a single-field model, there are two background variables populating the phase-space, $\phi$ and $\pi\equiv \mathrm{d}\phi/\mathrm{d}N$. 

In order to define the curvature perturbation $\zeta(t)$, let us introduce the number of e-folds elapsed between an initial time $t_i$ and some later time $t$, $N^{(t_i,t)}=\int_{t_i}^{t} \mathrm{d}t' \, H(t')$. 
By choosing at $t_i$ an initial spatially-flat hypersurface and a final one with constant energy density at $t$, $\zeta(t)$ is identified with the variation in the duration of inflation, $\delta N$, measured in each patch relative to the mean in a larger volume.
While this statement holds non-perturbatively, $\zeta(t)$ can be expanded in powers of phase-space perturbations, leading to the so-called $\delta N$ formula
\begin{equation}
\begin{split}
\label{eq: delta N formula}
    \zeta_\mathbf{k}(t) 
    &= 
    N_I^{(t_i,t)} \delta X^I_\mathbf{k}(t_i) 
    +
    \frac{1}{2!} N_{IJ}^{(t_i,t)} \int_{k_1\lesssim k_i} \mathrm{d}^3 k_1 
    \, \delta X^I_\mathbf{k_1} (t_i) 
    \, \delta X^J_\mathbf{k-k_1} (t_i) 
    \\
    &+ 
    \frac{1}{3!} N_{IJK}^{(t_i, t)}
    \int_{k_1\lesssim k_i} {\mathrm{d}^3 k_1}
    \int_{k_2\lesssim k_i} {\mathrm{d}^3 k_2}
    \, \delta X^I_{\mathbf{k_1}}(t_i)
    \, \delta X^J_{\mathbf{k_2}}(t_i)
    \, \delta X^K_{\mathbf{k}-\mathbf{k_1}-\mathbf{k_2}}(t_i)
    + \mathcal{O}(\delta X^4) \;.
\end{split}
\end{equation}
Here, upper-case latin indices label phase-space
coordinates and $\delta X^I_\mathbf{k}(t)$ represent phase-space perturbations. 
The $\delta N$ coefficients $N_I, \, N_{IJ}, \, \cdots$ measure the variation in the duration of inflation with respect to displacements in the phase-space background fields. 
Note that non-linear terms contain integrals which include the effect of modes larger than the separate universe smoothing scale, i.e. the one that crossed the horizon at the time of initialisation $(k_i\equiv a H|_{t_i})$.  

The choice of $t_i$ is at the heart of a correct applicatioon of the separate universe framework, as it must guarantee that any type of gradient correction to the evolution of perturbations can be neglected.  
In standard slow-roll models, this is ensured as long as Eq.~\eqref{eq: delta N formula} is initialised soon after horizon crossing, i.e. $t_i \gtrsim t_k$ where $k \equiv a H |_{t_k} $.
In models leading to a short-scale peak in $\mathcal{P}_\zeta(k)_\text{tree}$ the background features a sharp transition between slow-roll and a non-attractor phase.
These models are therefore characterised by a new scale, $k_\text{tr}$, and in order to correctly apply separate universe one has to wait for \textit{all} relevant scales to cross the horizon, including $k_\text{tr}$. 
In other words, $t_i$ must be chosen \textit{at a minimum} after the transition into the non-attractor phase has taken place~\cite{Jackson:2023obv}. 
Later on we will discuss the role played by this constraint in the computation of large-scale back-reaction. 
\vspace{0.3cm}

\noindent
\textbf{Properties of long and short modes.}
Before proceeding further, let us review the physical properties of the small- and large-scale modes. 
Our separate universe computation only relies on two fundamental assumptions: (i) the long mode is adiabatic, i.e. $\zeta_\mathbf{p}=\delta \phi_\mathbf{p}/\sqrt{2\epsilon_1}$ is frozen shortly after horizon crossing, and (ii) there is appreciable separation of scales between $p$ and the enhanced modes, namely $p\ll q$ in the language of Fig.~\ref{fig:USR}.

An important consequence of the large separation between $p$ and the short scale $q$ is that, upon horizon exit, $q$ experiences the long mode $p$ as a constant shift to the background fields.  
At the level of the short-scale 2-point function, the \textit{reaction} of $q$ to $p$ can be modeled as   
\begin{equation}
\label{eq:reaction}
    \langle \delta X^I_\mathbf{q}  \delta X^J_\mathbf{-q} \rangle_{t_q} 
    =
    \langle \delta X^I_\mathbf{q}  \delta X^J_\mathbf{-q} \rangle_{t_q}\Big|_0 
    + 
    \frac{\langle \delta X^I_\mathbf{q}  \delta X^J_\mathbf{-q} \rangle_{t_q}}{\partial X^L(t_q)} \Big|_0 
    \, 
    \delta X^L_\mathbf{p}(t_q)
    + \mathcal{O} \left({\delta X^L}^2 \right)\;, 
\end{equation}
where the notation $|_0$ indicates evaluation in the absence of the long mode $\delta X^L_\mathbf{p}(t_q)$. 
In this language the long-mode power spectrum at 1-loop can be seen as the \textit{back-reaction} of $p$ to $q$. 

The same operator product expansion used in Eq.~\eqref{eq:reaction} controls all long-short couplings, manifested in the form of squeezed correlators. 
For example, the reaction of $q$ to $p$ leads to the squeezed 3-point function~\cite{Maldacena:2002vr}
\begin{equation}
    B_{LIJ}(p,q,q;t_q)
    =
    P_{LM}(p,t_q)
    \frac{\partial P_{IJ}(q;t_q)}{\partial X^M(t_q)}\Big|_0 
    + \dots \;,
    \label{eq: squeezed bispectrum}
\end{equation}
where we have taken the limit $p\ll q$ and omitted corrections from higher powers of the long perturbation and from gradient corrections, the leading one being of order $(p/q)^2$. 
\vspace{0.3cm}

\noindent
\textbf{Long-mode back-reaction at 1-loop.}
As discussed below Eq.~\eqref{eq: delta N formula}, the non-linear terms in the $\delta N$ formula include the effect of modes larger than the smoothing scale, $k_i$. 
In order to set-up a $\delta N$ formula for $\zeta_\mathbf{p}(t)$ that includes the effect of all enhanced short scales, one must choose $t_i$ sufficiently late, such that all relevant amplified scales are super-horizon at $t_i$. 
Note that this criterion automatically ensures that separate universe is initialised after the transition into the non-attractor phase has taken place~\cite{Jackson:2023obv}. 
By using two copies of the $\delta N$ formula for $\zeta_\mathbf{p}(t)$ including back-reaction, one finds the following 1-loop contributions to the 2-point function~\cite{Iacconi:2023ggt,Iacconi:2026uzo}
\begin{equation}
    \langle \zeta_\mathbf{p}\zeta_\mathbf{-p} \rangle_\text{1-loop} 
    =
    \underbrace{\langle \zeta_\mathbf{p}\zeta_\mathbf{-p} \rangle_\text{11}}_{\text{11-type loops}}
    +
    \underbrace{\langle \zeta_\mathbf{p}\zeta_\mathbf{-p} \rangle_\text{22} + \langle \zeta_\mathbf{p}\zeta_\mathbf{-p} \rangle_\text{12} + \langle \zeta_\mathbf{p}\zeta_\mathbf{-p} \rangle_\text{13}}_{\delta N\text{ loops}} \;, 
\end{equation}
where each term has been labeled according to its building blocks from the $\delta N$ formula, e.g. the 12-type loop is constructed from the correlation between a linear and a quadratic term.
In a diagrammatic form, these are explicitly
\begin{equation}
\label{eq: 1-loop diagrams}
\begin{split}
\langle \zeta_\mathbf{p}\zeta_\mathbf{-p} \rangle_\text{11}&=
\eqgraph{1ex}{0ex}{%
\begin{fmfgraph*}(90,70)
  \fmfleft{l}
  \fmfright{r}
  \fmf{plain, label=$\zeta_\mathbf{p}$}{l,v1}
  \fmf{dashes}{v1,c1}
  \fmf{dashes,left,tension=0.4}{c1,c2}
  \fmf{dashes,left,tension=0.4}{c2,c1}
  \fmf{dashes}{c2,v2}
  \fmf{plain, label=$\zeta_\mathbf{-p}$}{v2,r}
  \fmfdot{v1,v2}
  \fmfv{label=$N_I$,label.angle=90,label.dist=4}{v1}
  \fmfv{label=$N_J$,label.angle=90,label.dist=4}{v2}
\end{fmfgraph*}}
\mspace{10mu}
+
\mspace{10mu}
\eqgraph{1ex}{0ex}{%
\begin{fmfgraph*}(90,70)
  \fmfleft{l}
  \fmfright{r}
  \fmf{plain, label=$\zeta_\mathbf{p}$, l.side=right}{l,v1}
  \fmf{dashes}{v1,c}
  \fmf{dashes, tension=0.6}{c,c}
  \fmf{dashes}{c,v2}
  \fmf{plain, label=$\zeta_\mathbf{-p}$}{v2,r}
  \fmfdot{v1,v2}
  \fmfv{label=$N_I$,label.angle=90,label.dist=4}{v1}
  \fmfv{label=$N_J$,label.angle=90,label.dist=4}{v2}
\end{fmfgraph*}}
\;, 
\\
\langle \zeta_\mathbf{p}\zeta_\mathbf{-p} \rangle_{\delta N}&= 
\eqgraph{1ex}{0ex}{%
\begin{fmfgraph*}(90,70)
  \fmfleft{l}
  \fmfright{r}
  \fmf{plain, label=$\zeta_\mathbf{p}$}{l,v1}
  \fmf{dashes,left,tension=0.4}{v1,v2}
  \fmf{dashes,left,tension=0.4}{v2,v1}
  \fmf{plain, label=$\zeta_\mathbf{-p}$}{v2,r}
  \fmfdot{v1,v2}
  \fmfv{label=$N_{IJ}$,label.angle=130,label.dist=4}{v1}
  \fmfv{label=$N_{KL}$,label.angle=50,label.dist=4}{v2}
\end{fmfgraph*}}
\mspace{10mu}
+
\mspace{10mu}
\eqgraph{1ex}{0ex}{%
\begin{fmfgraph*}(90,70)
  \fmfleft{l}
  \fmfright{r}
  \fmf{plain, label=$\zeta_\mathbf{p}$}{l,v1}
  \fmf{dashes}{v1,c}
  \fmf{dashes,left,tension=0.4}{c,v2}
  \fmf{dashes,left,tension=0.4}{v2,c}
  \fmf{plain, label=$\zeta_\mathbf{-p}$}{v2,r}
  \fmfdot{v1,v2}
  \fmfv{label=$N_I$,label.angle=90,label.dist=4}{v1}
  \fmfv{label=$N_{JK}$,label.angle=50,label.dist=4}{v2}
\end{fmfgraph*}}
\mspace{10mu}
+
\mspace{10mu}
\eqgraph{1ex}{0ex}{%
\begin{fmfgraph*}(90,70)
  \fmfleft{l}
  \fmfright{r}
  \fmf{plain, label=$\zeta_\mathbf{p}$}{l,v1}
  \fmf{dashes}{v1,v2}
  \fmf{dashes,right,tension=0.7}{v2,v2}
  \fmf{plain, label=$\zeta_\mathbf{-p}$}{v2,r}
  \fmfdot{v1,v2}
  \fmfv{label=$N_I$,label.angle=90,label.dist=4}{v1}
  \fmfv{label=$N_{JKL}$,label.angle=25,label.dist=6}{v2}
\end{fmfgraph*}}
\;.  
\end{split}
\end{equation}
Here, continuous lines represent $\zeta_\mathbf{p}(t)$, dashed lines mark correlators of phase-space initial conditions at $t_i$, and point-vertices embody $\delta N$ coefficients.   

Tackling the long-mode back-reaction computation within separate universe provides an easily interpretable result. 
Eq.~\eqref{eq: 1-loop diagrams} is organised into two types of contribution. 
The first line represents diagrams sourced by phase-space initial conditions at 1-loop, linearly evolved by separate universe from $t_i$ to $t$. 
The second line encodes diagrams where tree-level initial conditions in phase space are non-linearly evolved from $t_i$ to $t$ by separate universe.  
In this case, the 1-loop topology is due to non-linear evolution outside of the horizon. 
\vspace{0.3cm}

\noindent
\textbf{$\bm{\delta N}$ loops.}
The 22-type $\delta N$ loop is volume-suppressed by the separation of scales between long and short modes, $\langle\zeta \zeta\rangle_{22}\propto(p/q)^3$. 
It embodies the effect of Gaussian, uncorrelated noise on scale $q^{-1}$ averaged over a much larger region of size $p^{-1}$, with the scaling $(p/q)^3$ appearing as a consequence of the central limit theorem. 
From now on, we will neglect volume-suppressed contributions. 
On the other hand, the 12- and 13-type loops encode long-short mode couplings in the form of squeezed phase-space initial conditions, i.e. a genuine reaction of the short-scale statistics due to the long mode. 
Due to this feature, they evade volume suppression. 
\vspace{0.3 cm}

\noindent
\textbf{11-type loops and multi-point propagators.}
Diagrams with 11-type topology embody back-reaction that can be generated from the evolution of phase-space initial conditions from the first slow-roll period, through the non-attractor phase, up to $t_i$.
To compute these 1-loop phase-space correlators --represented with dashed lines in the first line of Eq.~\eqref{eq: 1-loop diagrams}-- one should find an appropriate expansion scheme which admits \textit{tree-level} initial conditions. 
This can be achieved by anchoring the expansion during the first slow-roll period. 
An initialisation time preceding the transition into the non-attractor phase is not compatible with separate universe~\cite{Jackson:2023obv}.
Instead, one can employ a ``multi-point propagator'' (MPP) expansion~\cite{Costantini:2025tek}
\begin{equation}
\begin{split}
\label{eq: MPP expansion}
    \delta X^I_\mathbf{k}(t_i) 
    &= 
    {\Gamma^I_J}(k)^{(t_\star,t_i)} \delta X^J_\mathbf{k}(t_\star) 
    +
    \frac{1}{2!} {\Gamma^I_{JK}}(k,k_1,|\mathbf{k-k_1}|)^{(t_\star,t_i)} \int \mathrm{d}^3 k_1 
    \, 
    \delta X^J_\mathbf{k_1} (t_\star) 
    \, 
    \delta X^K_\mathbf{k-k_1} (t_\star) 
    \\
    &+ 
    \frac{1}{3!} {\Gamma^I_{JKL}} (\mathbf{k}, \mathbf{k_1},\mathbf{k_2}, \mathbf{k}-\mathbf{k_1}-\mathbf{k_2})^{(t_\star, t_i)}
    \int {\mathrm{d}^3 k_1}
    \int {\mathrm{d}^3 k_2}
    \,
    \delta X^J_{\mathbf{k_1}}(t_\star) 
    \,
    \delta X^K_{\mathbf{k_2}}(t_\star)
    \,
    \delta X^L_{\mathbf{k}-\mathbf{k_1}-\mathbf{k_2}}(t_\star)
    \\
    &+ \mathcal{O}(\delta X^4) \;,
\end{split}
\end{equation}
where we have labeled the base time as $t_\star$.
Note that since there is no smoothing scale in the expansion, the integrals are not limited to specific scales. 
The MPP framework is more general, does not rely on validity of the separate universe picture and therefore allows for more flexibility. 
Nevertheless, the $\Gamma$ coefficients are scale-dependent objects, and their interpretation is less straightforward than that of $\delta N$ coefficients. 

\section{Decoupling at 1-loop between large and small scales}

In order to tackle the fundamental question of whether back-reaction is measured by the amplitude of the short-scale peak, it is useful to investigate the sum of 11-type loops and (separately) of 12- and 13-type $\delta N$ loops. 
\vspace{0.3cm}

\noindent
\textbf{Boundary contributions.}
Under the assumptions of adiabaticity of the long mode and separation of scales, one finds~\cite{Iacconi:2026uzo}
\begin{equation}
    \label{eq: results}
    \begin{split}
    \mathcal{P}_\zeta(p)_\text{11-type loops}
    &\propto
    \int_{q_\text{min}}^{q_\text{max}} 
    \mathrm{d} \ln q \,
    \frac{\mathrm{d}}{\mathrm{d} \ln q}
    \left[
    {\Gamma^J_{KM}}(p,q,q)^{(t_\star,t_i)}\, \mathcal{P}^{KM}(q;t_\star)_{\text{tree}}
    \right] \;,
    \\
    \mathcal{P}_\zeta(p)_\text{12+13-type loops}
    &\propto
    \int_{q_\text{min}}^{q_\text{max}} 
    \mathrm{d} \ln q \,
    \frac{\mathrm{d}}{\mathrm{d} \ln q}
    \left[
    N_{JK}^{(t_i,t)}\, \mathcal{P}^{JK}(q;t_i)_{\text{tree}}
    \right] \;. 
    \end{split}
\end{equation}
In both cases, the emergence of a total derivative is a direct consequence of adiabaticity of the long mode. 
The main conclusion to draw from Eq.~\eqref{eq: results} is that 
non-volume-suppressed back-reaction reduces to contributions from the integral boundaries only.
Interestingly, other authors also obtained similar results in the context of In--In computations~\cite{Tada:2023rgp,Fumagalli:2023zzl,Kawaguchi:2024rsv}. 

Evaluation of Eq.~\eqref{eq: results} is degenerate with the (seemingly arbitrary) choice made for $q_\text{min}$ and $q_\text{max}$. 
Since our main objective is to isolate the effect of the short-scale peak, we identify $q_\text{min}$ and $q_\text{max}$ with a scale preceding and one following it respectively. 
In this way, all modes that populate the peak are included in the integrals in Eq.~\eqref{eq: results}.
Note that this choice also minimises cancellation with adjacent momentum regions, e.g. $q\lesssim q_\text{min}$ and $q\gtrsim q_\text{max}$, that one must take into account in the full 1-loop computation. 

The back-reaction at 1-loop in Eq.~\eqref{eq: results} is model-dependent, and crucially it is scale-invariant.
It is therefore degenerate with effects from UV modes that were still in the vacuum at the end of inflation. 
We conclude that the effect found is not observable.
In this sense, an adiabatic long mode decouples at 1-loop from enhanced perturbations on small scales. 
\vspace{0.3cm}

\begin{figure}
\begin{minipage}{0.4\linewidth}
\centerline{\includegraphics[width=\linewidth]{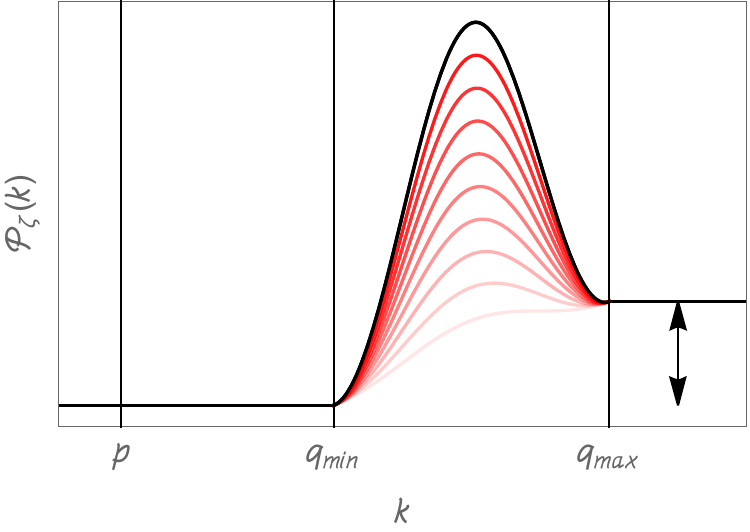}}
\end{minipage}
\hfill
\begin{minipage}{0.4\linewidth}
\centerline{\includegraphics[width=\linewidth]{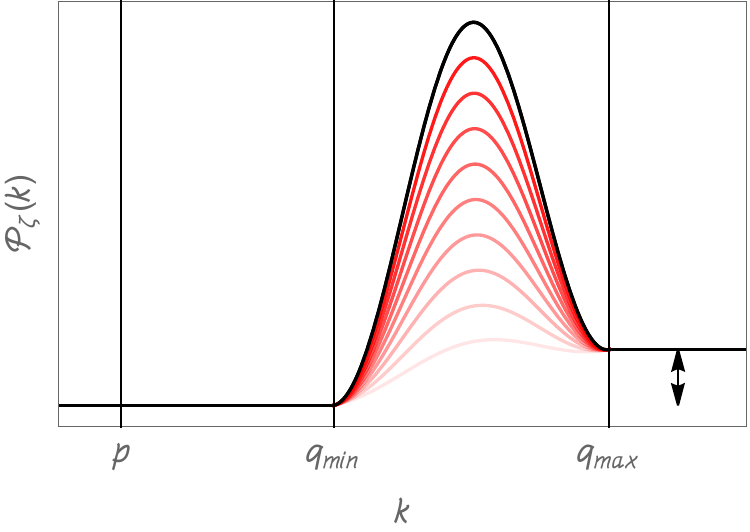}}
\end{minipage}
\caption[]{Schematic illustration of tree-level power spectra displaying a peak on short scales.
These are not computed from realistic inflationary models (such as those in Fig.~\ref{fig:USR}), but
rather are toy examples. 
The vertical, black arrow indicates the (model-dependent) quantity $\mathcal{P}_\zeta(q_\text{max};t)_\text{tree} - \mathcal{P}_\zeta(q_\text{min};t)_\text{tree}$. In each panel, a reference spectrum is plotted in black. 
The red lines represent examples of other spectra with the same infrared and ultraviolet plateaus
as the reference one, but with different amplitude of the peak.}
\label{fig: illustration of results}
\end{figure}

\noindent
\textbf{An example with ultra-slow-roll.}
For the case of a model featuring a USR phase, it is possible to give a more familiar expression for the back-reaction due to $\delta N$ loops. 
By initialising separate universe in the last slow-roll phase (when $\epsilon_1(t_i), \epsilon_2(t_i) \ll1$), most of the enhanced modes are included in the loop integral. 
Moreover, since $t_i$ is chosen during an attractor phase velocity perturbations can be neglected. 
In this way, the $\delta N$ loops in Eq.~\eqref{eq: results} simplify to 
\begin{equation}
    \label{eq: USR results}
    \mathcal{P}_\zeta(p)_\text{12+13-type loops} 
    \propto 
    f^{N_{\phi\phi}}_\text{NL,eq}(k_i;t) 
    \Big[ 
    \mathcal{P}_\zeta(q_\text{max};t)_\text{tree}
    -
    \mathcal{P}_\zeta(q_\text{min};t)_\text{tree}
    \Big]  \;,
\end{equation}
where $f^{N_{\phi\phi}}_\text{NL,eq}$ is the amplitude of equilateral non-Gaussianity generated on super-horizon scales. 
Clearly the result does not depend on detailed properties of the small-scale peak itself. 
In particular, it is not controlled by its amplitude, $\mathcal{P}_\zeta(k_\text{peak})$, which in turns determines PBHs abundance.
See Fig.~\ref{fig: illustration of results} for an illustration of the meaning of Eq.~\eqref{eq: USR results}. 
Therefore, the existence of a PBH-producing peak does not alter the statistical properties on large scales. 

\section{Summary and new directions}

PBHs are significant in cosmology for many reasons~\cite{Carr:2026hot}. 
They also provide a unique window for testing inflation on scales much smaller than those probed by CMB experiments. 
PBHs can be produced from inflation from the collapse of large and rare primordial fluctuations. 
The substantial enhancement of small-scale power required for PBH production can be realised within single-field models by means of a transient non-attractor phase in between slow-roll eras. 
In the presence of long-short mode couplings, enhanced short scales can induce non-volume-suppressed back-reaction of long modes~\cite{Iacconi:2023ggt}.
In a loop expansion of the long-mode power spectrum, this
back-reaction appears first at 1-loop. 
Nevertheless, \textit{tree-level} slow-roll predictions are in excellent agreement with CMB observations, and observable back-reaction at \textit{1-loop} from the enhanced modes could potentially alter the statistics on large scales. 
In these proceedings we have reviewed the separate universe approach to the computation of large-scale back-reaction~\cite{Iacconi:2023ggt,Iacconi:2026uzo}. 
We rely on assumption of (i) adiabaticity of the long mode, and (ii) large separation between long and short scales. 
We find two types of non-volume-suppressed contributions to back-reaction: 11-type loops, sourced by initial conditions at 1-loop, and $\delta N$-type loops, which are due to non-linear super-horizon evolution. 
By considering the collective effect of a realistic, localised peak on short scales, we find that both types of contributions can be organised as a momentum integral of a total derivative, thereby indicating that the short-scale modes only contribute at the boundaries of the loop integral. 
In particular, the back-reaction is not measured from the detailed properties of the peak, including its amplitude. 
Our results show that an adiabatic long mode decouples from the collective effect of enhanced short scales at 1-loop. 

While back-reaction in this scenario is unobservable, this is not necessarily true in other cases. 
For example, our result does not apply to ``self-corrections'' of the peak scales. 
Indeed, this computation does not feature a hierarchy of scales and the short modes are not adiabatic already at tree-level. 
Since PBH formation is exponentially sensitive to the characteristics of the short scale peak, the investigation of peak scales self-corrections is clearly worthwhile. 
Our result also does not apply straightforwardly to situations in which the long mode itself is non-adiabatic, as it is usually the case within multi-field models. 
For scenarios with observable back-reaction, it would be possible to employ large-scale measurements to constrain inflation on much smaller scales.

\section*{Acknowledgments}

I thank the organizers of the 2026 Cosmology session of the 60th Rencontres de Moriond for their invitation, and the participants for many stimulating discussions. 
These proceedings are based on work done in collaboration with David Mulryne and David Seery, and supported by the Science and Technology Facilities Council
(grant number ST/X000931/1). 

\bibliography{iacconilaura}

\end{fmffile}

\end{document}